\documentclass[journal]{IEEEtran}
\ifCLASSINFOpdf
\else
   \usepackage[dvips]{graphicx}
\fi
\usepackage{url}
\usepackage{graphicx}
\usepackage{times}
\usepackage{epsfig}
\usepackage{graphicx}
\usepackage{amsmath}
\usepackage{amssymb}
\usepackage{longtable}
\usepackage{amsfonts}
\usepackage{algorithmic}
\usepackage{algorithm}
\usepackage{color}
\usepackage{url}
\usepackage{tabularx}
\usepackage{array,multirow}
\usepackage{amssymb}
\usepackage{caption}
\usepackage{subcaption}
\usepackage[outdir=./]{epstopdf}
\DeclareMathSymbol{\shortminus}{\mathbin}{AMSa}{"39}

\begin{document}
\title{Variable Rate Deep Image Compression with Modulated Autoencoder}
\author{Fei Yang, Luis Herranz, Joost van de Weijer, José A. Iglesias Guitián, Antonio M. López, Mikhail G. Mozerov

\thanks{All authors are with the Computer Vision Center at the Universitat Autònoma de Barcelona (UAB). Antonio and Joost are also with the Computer Science Dpt. at UAB. Fei is also with Key Laboratory of Information Fusion Technology, Northwestern Polytechnical University, China. E-mail: fyang@cvc.uab.es}

\thanks{The authors thank Audi Electronics Venture GmbH for supporting this work, the Generalitat de Catalunya CERCA Program and its ACCIO agency. Fei Yang acknowledges the Chinese Scholarship Council grant No.201706290127. Luis acknowledges the support of the Spanish project RTI2018-102285-A-I00. Joost acknowledges the support of the Spanish project TIN2016-79717-R. Antonio and Jose acknowledge the support of project TIN2017-88709-R (MINECO/AEI/FEDER, UE). Antonio also thanks the support by ICREA Academia programme. Jose and Luis acknowledge the support of the EU’s Horizon 2020 R\&I programme under the Marie Skłodowska-Curie grant No.665919.}}

\markboth{Journal of \LaTeX\ Class Files, Vol. 14, No. 8, August 2019}
{Shell \MakeLowercase{\textit{et al.}}: Bare Demo of IEEEtran.cls for IEEE Journals}
\maketitle

\begin{abstract}
Variable rate is a requirement for flexible and adaptable image and video compression. However, deep image compression methods (DIC) are optimized for a single fixed rate-distortion (R-D) tradeoff. While this can be addressed by training multiple models for different tradeoffs, the memory requirements increase proportionally to the number of models. Scaling the bottleneck representation of a shared autoencoder can provide variable rate compression with a single shared autoencoder. However, the R-D performance using this simple mechanism degrades in low bitrates, and also shrinks the effective range of bitrates.
To address these limitations, we formulate the problem of variable R-D optimization for DIC, and propose modulated autoencoders (MAEs), where the representations  of a shared autoencoder are adapted to the specific R-D tradeoff via a modulation network. Jointly training this modulated autoencoder and the modulation network provides an effective way to navigate the R-D operational curve. Our experiments show that the proposed method can achieve almost the same R-D performance of independent models with significantly fewer parameters.
\end{abstract}

\begin{IEEEkeywords}
Deep image compression, variable bitrate, autoencoder, modulated autoencoder
\end{IEEEkeywords}

\IEEEpeerreviewmaketitle

\section{Introduction}
\begin{figure}[!t]
	\centering
	\includegraphics[width=\columnwidth]{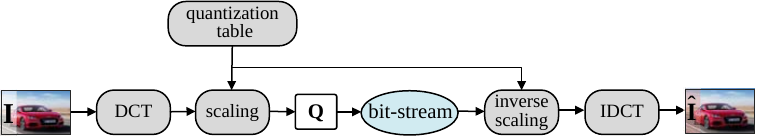}\\
	(a)\\
	\vspace{0.1cm}
	\includegraphics[width=\columnwidth]{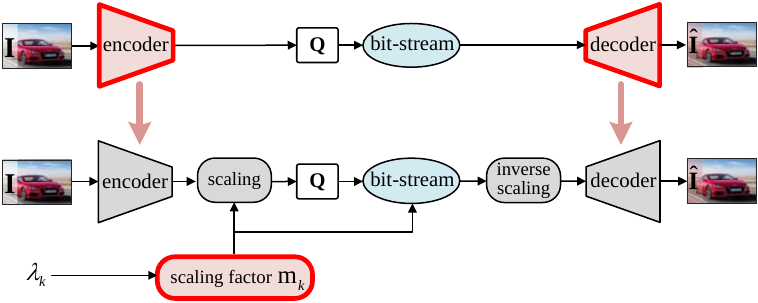}\\
	(b)\\
	\vspace{0.1cm}
	\includegraphics[width=\columnwidth]{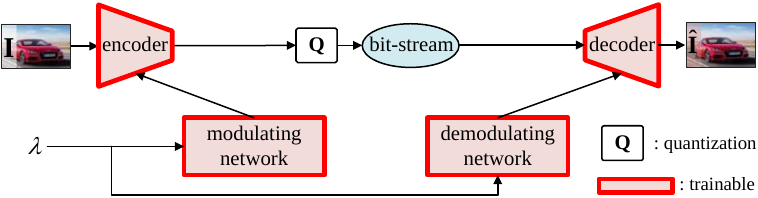}\\
	(c)\\
	\caption{Image compression and R-D control: (a) JPEG transform, (b) pre-trained autoencoder with bottleneck scaling~\cite{theis2017lossy}, and (c) our proposed modulated autoencoder with joint training. Entropy coding/decoding is omitted for simplicity.}
	\label{fig:motivation_}
	\vspace{-0.5cm}
\end{figure}
\IEEEPARstart{I}{mage} compression is a fundamental and well-studied problem in image processing and computer vision~\cite{lewis1992image,villasenor1995wavelet,taubman2012jpeg2000}.  The goal is to design binary representations (i.e. bitstreams) with minimal entropy~\cite{shannon1948mathematical} that minimize the number of bits required to represent an image (i.e. bitrate) at a given level of fidelity (i.e. distortion)~\cite{jain1989fundamentals}. In many applications, communication networks or storage devices may impose a constraint on the maximum bitrate, which requires the image encoder to adapt to a given bitrate budget. In some applications this constraint may even change dynamically over time (e.g. video transmission). In all these cases, a bitrate control mechanism is required, and it is available in most traditional image and video compression codecs. In general, reducing the bitrate causes an increase in the distortion, i.e. there is a rate-distortion (R-D) tradeoff. This mechanism is typically based on scaling the latent representation prior to quantization to obtain finer or coarser quantizations, and then inverting the scaling at the decoder side (Fig.~\ref{fig:motivation_}a).

Recent studies show that deep image compression (DIC) achieves comparable or even better results than classical image compression techniques~\cite{toderici2015variable, balle2016end,gregor2016towards,toderici2017full,theis2017lossy,johnston2018improved, liu2018cnn,mentzer2018conditional, minnen2018joint,li2018learning}. In this paradigm, the parameters of the encoder and decoder are learned from certain image data by jointly minimizing rate and distortion at a particular R-D tradeoff. However, variable bitrate requires an independent model for every R-D tradeoff. This is an obvious limitation, since it requires storing each model separately, resulting in large memory requirement. 

To address this limitation, Theis et al.~\cite{theis2017lossy} use a single autoencoder whose bottleneck representation is scaled before quantization depending on the target bitrate (Fig.~\ref{fig:motivation_}b). However, this approach only considers the importance of different channels from the bottleneck representation of learned autoencoders under R-D tradeoff constraint. In addition, the autoencoder is optimized for a single specific R-D tradeoff (typically high bitrate). These aspects lead to a drop in performance for low bitrates and a narrow effective range of bitrates.

In order to tackle the limitations of multiple independent models and bottleneck scaling, we formulate the problem of variable R-D optimization for DIC, and propose the \textit{modulated autoencoder} (MAE) framework, where the representations of a shared autoencoder at different layers are adapted to a specific R-D tradeoff via a modulating network. The modulating network is conditioned on the target R-D tradeoff, and synchronized with the actual tradeoff optimized to learn the parameters of the autoencoder and the modulating network. MAEs can achieve almost the same operational R-D points of independent models with much fewer overall parameters (i.e. just the shared autoencoder plus the small overhead of the modulating network). Multi-layer modulation does not suffer from the main limitations of bottleneck scaling, namely, drop in performance for low rates, and shrinkage of the effective range of bitrates.

\section{background}
Almost all lossy image and video compression approaches follow the transform coding paradigm~\cite{goyal2001theoretical}. The basic structure is a transform $\mathbf{z}=f\left(\mathbf{x}\right)$ that takes an input image $\mathbf{x}\in \mathbb{R}^{N}$ and obtains a transformed representation $\mathbf{z}$, followed by a quantizer $\mathbf{q}=Q\left(\mathbf{z}\right)$ where $\mathbf{q}\in \mathbb{Z}^D$ is a discrete-valued vector. The decoder reverses the quantization (i.e. dequantizer $\mathbf{\hat{z}}=Q^{-1}\left(\mathbf{q}\right)$) and the transform (i.e. inverse transform) as $\mathbf{\hat{x}}=g\left(\mathbf{\hat{z}}\right)$ reconstructing the output image $\mathbf{\hat{x}}\in \mathbb{R}^N$. Before the transmission (or storage), the discrete-valued vector $\mathbf{q}$ is binarized and serialized into a \textit{bitstream} $\mathbf{b}$. Entropy coding~\cite{wintz1972transform} is used to exploit the statistical redundancy in that bitstream and reduce its length.

In DIC, the handcrafted analysis and synthesis transforms are replaced by the encoder $\mathbf{z}=f\left(\mathbf{x};\theta\right)$ and decoder $\mathbf{\hat{x}}=g\left(\mathbf{\hat{z}};\phi\right)$ of a convolutional autoencoder, parametrized by $\theta$ and $\phi$. The fundamental difference is that the transforms are not designed but \textit{learned} from training data. 
The model is typically trained by minimizing the optimization problem
\begin{equation}
	{\arg \min}_{\theta,\phi} R\left(\mathbf{b}\right) + \lambda D\left(\mathbf{x},\mathbf{\hat{x}}\right),
	\label{eq:RDproblem}
\end{equation}
where $R\left(\mathbf{b}\right)$ measures the rate of the bitstream $\mathbf{b}$ and $D\left(\mathbf{\hat{x}},\mathbf{x}\right)$ represents a distortion metric between $\mathbf{x}$ and $\hat{\mathbf{x}}$, and the Lagrange multiplier $\lambda$ controls the R-D tradeoff. Note that $\lambda$ is fixed in this case. The problem is solved using gradient descent and backpropagation~\cite{rumelhart1985learning}.

To make the model differentiable, which is required to apply backpropagation,  during training the quantizer is replaced by a differentiable proxy function~\cite{theis2017lossy,toderici2015variable,balle2016end}. Similarly, entropy coding is invertible, but it is necessary to compute the length of the bitstream $\mathbf{b}$. This is usually approximated by the entropy of the distribution of the quantized vector, $R\left(\mathbf{b}\right)\approx H\left[P_\mathbf{q}\right]$, which is a lower bound of the actual bitstream length.

In this paper, we will use scalar quantization by (element-wise) rounding to the nearest neighboor, i.e. $\mathbf{q}=\lfloor\mathbf{z}\rfloor$, which will be replaced by additive uniform noise as proxy during training, i.e. $\mathbf{\tilde{z}}=\mathbf{z}+\Delta \mathbf{z}$, with $\Delta \mathbf{z}\sim \mathcal{U}\left(-\frac{1}{2},\frac{1}{2}\right)$. There is no de-quantization in the decoder, and the reconstructed representation is simply $\mathbf{\hat{z}}=\mathbf{q}$. To estimate the entropy we will use the entropy model described in~\cite{balle2016end} to approximate $P_\mathbf{q}$ by $p_\mathbf{\tilde{z}}\left(\mathbf{\tilde{z}}\right)$. Finally, we will use  mean squared error (MSE) as a distortion metric. With these particular choices, (\ref{eq:RDproblem}) becomes
\begin{equation}
	\operatornamewithlimits{argmin}\limits_{\theta,\phi} R\left(\mathbf{\tilde{z}};\theta\right)
	+ \lambda D\left(\mathbf{x},\mathbf{\hat{x}};\theta,\phi\right),
	\label{eq:RDproblem2}
\end{equation}
with
\begin{align}
    R\left(\mathbf{\tilde{z}};\theta\right) &= \mathbb{E}_{\mathbf{x}\sim p_\mathbf{x},\Delta\mathbf{z}\sim \mathcal{U}}\left[
	-\log_2 p_\mathbf{\tilde{z}}\left(\mathbf{\tilde{z}}\right)\right],
    \\
    D\left(\mathbf{x},\mathbf{\hat{x}};\theta,\phi\right) &= \mathbb{E}_{\mathbf{x}\sim p_\mathbf{x},\Delta\mathbf{z}\sim \mathcal{U}}\left[\left\|\mathbf{x}-\mathbf{\hat{x}}\right\|^2\right].
\end{align}
\begin{figure*}[!t]
  \centering
  \includegraphics[width=0.98\textwidth]{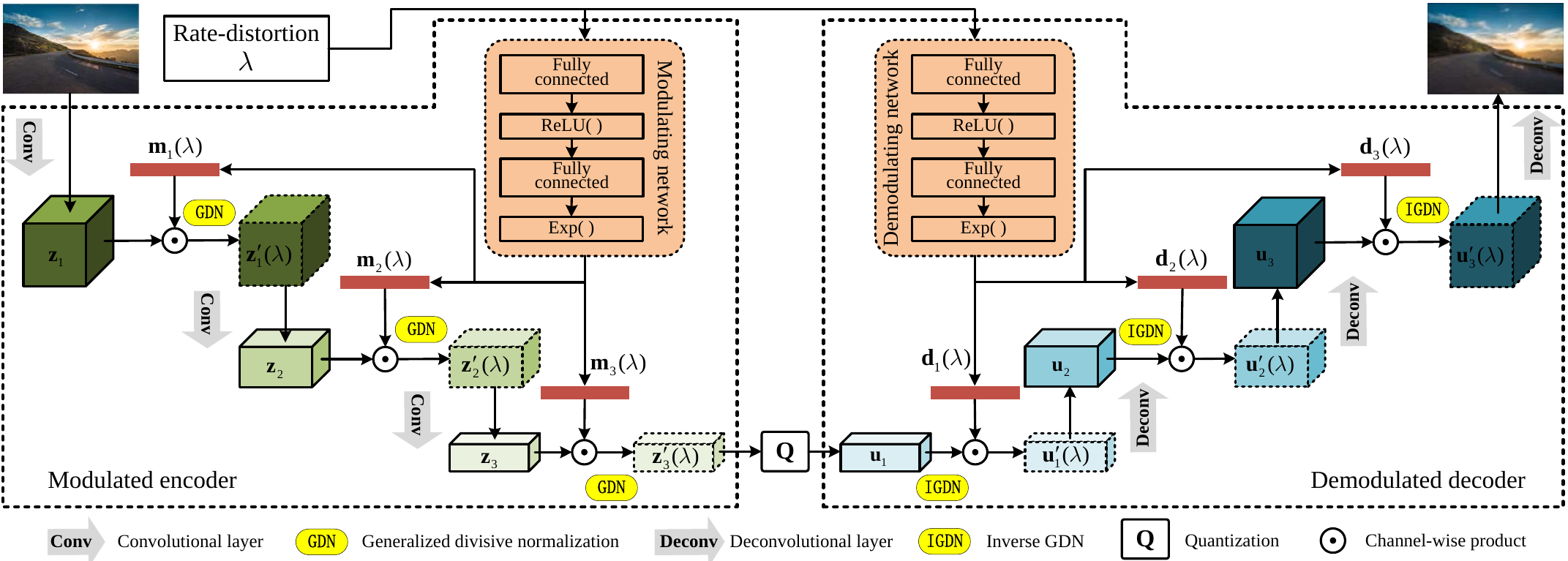}
  \caption{Modulated autoencoder (MAE) architecture, combining modulating networks and shared autoencoder. The channel-wise product is performed before GDN in the encoder and after IGDN in the decoder.
  }
  \label{MAE}
  \vspace{-0.4cm}
\end{figure*}
\vspace{-0.5cm}
\section{Multi-rate deep image compression with modulated autoencoders}
\subsection{Problem definition}
We are interested in DIC models that can operate satisfactorily on different R-D tradeoffs, and adapt to a specific one when required. Note that Eq.(\ref{eq:RDproblem2}) optimizes rate and distortion for a fixed tradeoff $\lambda$. We extend that formulation to multiple R-D tradeoffs (i.e. $\lambda\in\Lambda=\left\{\lambda_1,\ldots,\lambda_M\right\}$) as the \textit{multi-rate-distortion problem}
\begin{equation}
	\operatornamewithlimits{argmin}\limits_{\theta,\phi} \sum_{\lambda\in \Lambda} \left[ R\left(\mathbf{\tilde{z}};\theta,\lambda\right)
	+ \lambda D\left(\mathbf{x},\mathbf{\hat{x}};\theta,\phi,\lambda\right)\right],
	\label{eq:discreteMRDproblem}
\end{equation}
with
\begin{align}
    R\left(\mathbf{\tilde{z}};\theta,\lambda\right) &= \mathbb{E}_{\mathbf{x}\sim p_\mathbf{x},\Delta\mathbf{z}\sim \mathcal{U}}\left[
	-\log_2 p_\mathbf{\tilde{z}}\left(\mathbf{\tilde{z}}\right)\right],
    \\
    D\left(\mathbf{x},\mathbf{\hat{x}};\theta,\phi,\lambda\right) &= \mathbb{E}_{\mathbf{x}\sim p_\mathbf{x},\Delta\mathbf{z}\sim \mathcal{U}}\left[\left\|\mathbf{x}-\mathbf{\hat{x}}\right\|^2\right],
\end{align}
where we are simplifying the notation by omitting features dependency on $\lambda$, i.e. $\mathbf{\tilde{z}}=\mathbf{\tilde{z}}\left(\lambda\right)=f\left(\mathbf{x};\theta,\lambda\right)$ and $\mathbf{\hat{x}}=\mathbf{\hat{x}}\left(\lambda\right)=g\left(\mathbf{\tilde{z}\left(\lambda\right)};\phi,\lambda\right)$. This formulation can be easily extended to a continuous range of tradeoffs. Note also that these optimization problems assume that all R-D operational points are equally important. It could be possible to integrate an importance function  $I\left(\lambda\right)$ to further give more importance to certain R-D operational points if required. We assume uniform importance (continuous or discrete) for simplicity.

\subsection{Bottleneck scaling}
A possible way to make the encoder and decoder aware of $\lambda$ is  simply scaling the latent representation in the bottleneck before quantization (implicitly scaling the quantization bin), and then inverting the scaling in the decoder. In this case, $\mathbf{q}=Q\left(\mathbf{z}\odot \mathbf{s}\left(\lambda\right)\right)$ and $\mathbf{\hat{x}}\left(\lambda\right)=g\left(\mathbf{\hat{z}\left(\lambda\right)\odot \left(1/\mathbf{s}\left(\lambda\right)\right)};\phi\right)$,  where $\mathbf{s}\left(\lambda\right)$ is the specific scaling factor for the tradeoff $\lambda$. Conventional codecs use predefined tables for $\mathbf{s}\left(\lambda\right)$ (the descaling is often implicitly subsumed in the dequantization, e.g. JPEG), while other approaches~\cite{theis2017lossy} keep encoder and decoder fixed, optimized for a particular R-D tradeoff (Fig.~\ref{fig:motivation_}(b)).

We observe several limitations in this approach: (1) scaling only the bottleneck features is not flexible enough to adapt to a large range of R-D tradeoffs, (2) using the inverse of the scaling factor in the decoder may also limit the flexibility of the adaptation mechanism, (3) optimizing the parameters of the autoencoder only for a single R-D tradeoff leads to suboptimal parameters for other distant tradeoffs, (4) training the autoencoder and the scaling factors separately may also be limiting. In order to overcome these limitations we propose the \textit{modulated autoencoder (MAE)} framework.
\vspace{-0.3cm}
\subsection{Modulated autoencoders}
Variable rate is achieved in MAEs by modulating the internal representations in the encoder and the decoder (Fig.~\ref{MAE}). Given a set of internal representations in the encoder $\mathbf{Z}=\left\{\mathbf{z}_1,\ldots,\mathbf{z}_K\right\}$ and in the decoder $\mathbf{U}=\left\{\mathbf{u}_1,\ldots,\mathbf{u}_L\right\}$, they are replaced by the corresponding modulated and demodulated versions $\mathbf{Z}'=\left\{\mathbf{z}_1\odot \mathbf{m}_1\left(\lambda\right),\ldots,\mathbf{z}_K\odot \mathbf{m}_K\left(\lambda\right)\right\}$ and $\mathbf{U}'=\left\{\mathbf{u}_1\odot \mathbf{d}_1\left(\lambda\right),\ldots,\mathbf{u}_L\odot \mathbf{d}_L\left(\lambda\right)\right\}$, where $\mathbf{m}\left(\lambda\right)=\left(\mathbf{m}_1\left(\lambda\right),\ldots,\mathbf{m}_K\left(\lambda\right)\right)$ and $\mathbf{d}\left(\lambda\right)=\left(\mathbf{d}_1\left(\lambda\right),\ldots,\mathbf{d}_L\left(\lambda\right)\right)$ are the modulating and demodulating functions. 

Our MAE architecture extends the DIC architecture proposed in \cite{balle2016end} which combines convolutional layers and GDN/IGDN layers~\cite{balle2015density}. In our experiments, we choose to modulate the outputs of the convolutional layers in the encoder and decoder, i.e.  $\mathbf{Z}$ and $\mathbf{U}$, respectively.

The modulating function $\mathbf{m}\left(\lambda\right)$ for the encoder is learned by a \textit{modulating network} 
as $\mathbf{m}\left(\lambda\right)=\mathbf{m}\left(\lambda;\vartheta\right)$ and the demodulating function $\mathbf{d}\left(\lambda\right)$ by the \textit{demodulating network} as $\mathbf{d}\left(\lambda\right)=d\left(\lambda;\varphi\right)$.
As a result, the encoder has learnable parameters $\left\{\theta,\vartheta\right\}$ and the decoder $\left\{\phi,\varphi\right\}$.

Finally, the optimization problem for the MAE is 
\begin{equation}
	\operatornamewithlimits{argmin}\limits_{\theta,\phi,\vartheta,\varphi} \sum_{\lambda\in \Lambda} \left[ R\left(\mathbf{\tilde{z}};\theta,\vartheta,\lambda\right)
	+ \lambda D\left(\mathbf{x},\mathbf{\hat{x}};\theta,\phi,\vartheta,\varphi,\lambda\right)\right],
	\label{eq:MAEproblem}
\end{equation}
which extends Eq.(\ref{eq:discreteMRDproblem}) with the modulating/demodulating networks and their corresponding parameters. All parameters are learned jointly using gradient descent and backpropagation.

This mechanism is more flexible than bottleneck scaling since it allows multi-level modulation, decouples encoder and decoder scaling and allows effective joint training of both autoencoder and modulating network, therefore optimizing jointly to all R-D tradeoffs of interest.

\subsection{Modulating and demodulating networks}
The modulating network is a perceptron with two FC layers and ReLU~\cite{nair2010rectified} and exponential nonlinearities (Fig.~\ref{MAE}). The exponential nonlinearity guarantees positive outputs which we found beneficial in training. The input is a scalar value $\lambda$ and the output is $\mathbf{m}\left(\lambda\right)=\left(\mathbf{m}_1\left(\lambda\right),\ldots,\mathbf{m}_K\left(\lambda\right)\right)$. A small first hidden layer allows learning a meaningful nonlinear function between tradeoffs and modulation vectors, which is more flexible than simple scaling factors and allows more expressive interpolation between tradeoffs. In practice, we use normalized tradeoffs as $\hat{\lambda}_k=\lambda_k/\text{max}_{\lambda\in \Lambda}\left(\lambda\right)$. The demodulating network follows a similar architecture.

\section{Experiments}
\subsection{Experimental setup}
We evaluated MAE on the CLIC 2019 Professional dataset with 585 training images and 226 test images. In addition, we also test our models on Kodak dataset.
Our implementation~\footnote{https://github.com/FireFYF/modulatedautoencoder} is based on the autoencoder architecture of \cite{balle2016end}, which is augmented with modulation mechanisms and modulating networks (two FC layers, with $150$ and $3\times192$ units respectively) for all the convolutional layers. We use MSE as distortion metric. The model is trained with crops of size $240\times240$ using Adam with a minibatch size of 8 and initial learning rates of 0.0004 and 0.002 for MAE and the entropy model, respectively. After 400k iterations, the learning rates are halved for another 150k iterations. We also tested MAEs with scale hyperpriors, as described in \cite{balle2018variational}. In our experiments, we consider seven ($\lambda \in [64, 128, 256, 512, 1024, 2048, 4096]$) and four ($\lambda \in [64, 256, 1024, 4096]$) R-D tradeoffs for the models without and with hyperprior, respectively. We consider two baselines:

\textbf{Independent models}. Each R-D operational point is obtained by training a new model with a different R-D tradeoff $\lambda$ in (\ref{eq:RDproblem2}), requiring each model to be stored separately. This provides the optimal R-D performance, but also requires more memory to store all the models for different R-D tradeoffs.

\textbf{Bottleneck scaling}. The autoencoder is optimized for the highest R-D tradeoff in the range. Then it is frozen and the scaling parameters are learned for the other tradeoffs. 
\vspace{-5pt}
\begin{figure}[!t]
  \centering
  \includegraphics[width=0.49\textwidth]{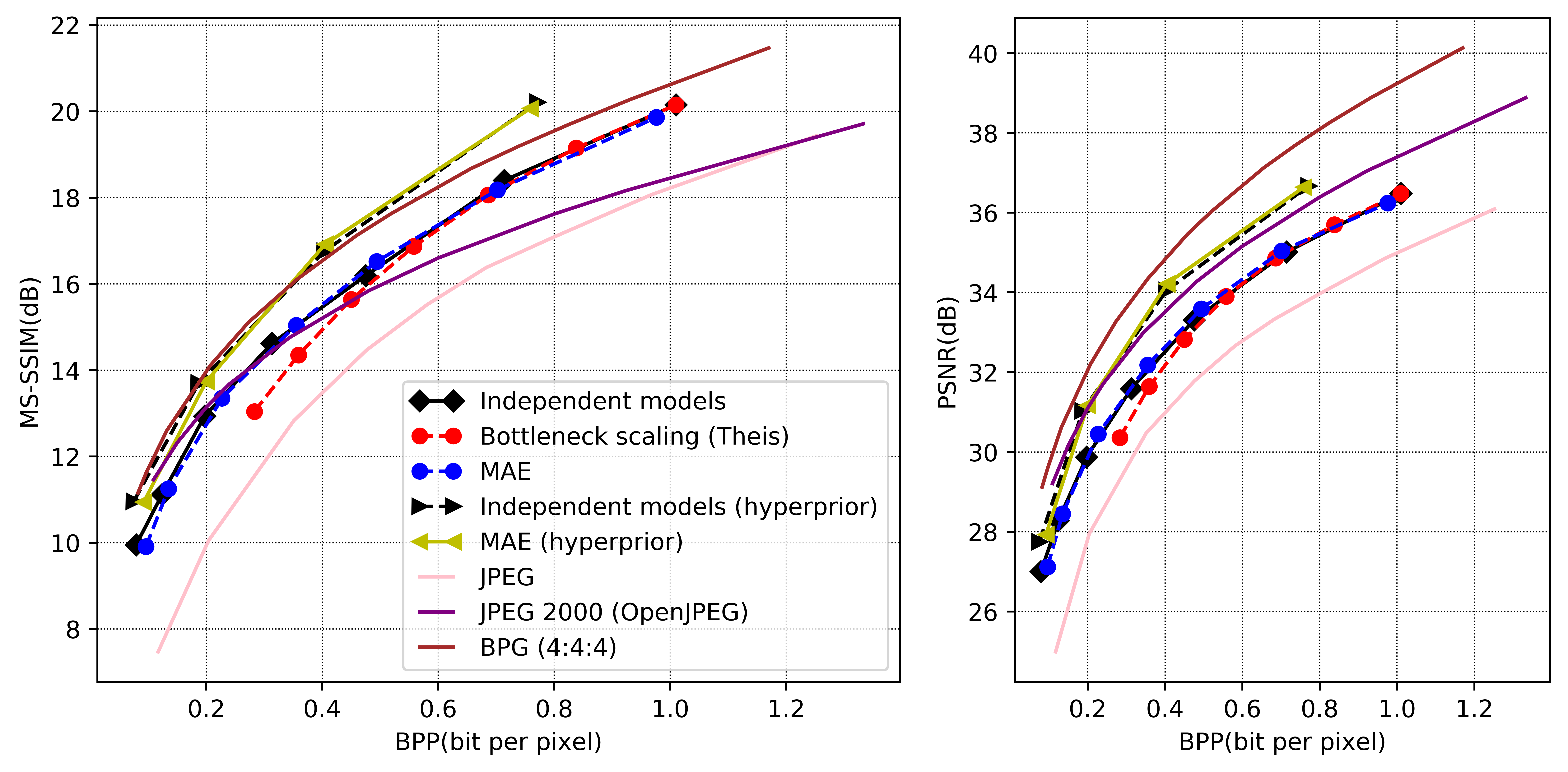}
  \caption{R-D curves for different methods on CLIC 2019 Professional test dataset.}
  \label{fig:clic}
\end{figure}
\begin{figure}[!t]
  \centering
  \includegraphics[width=0.49\textwidth]{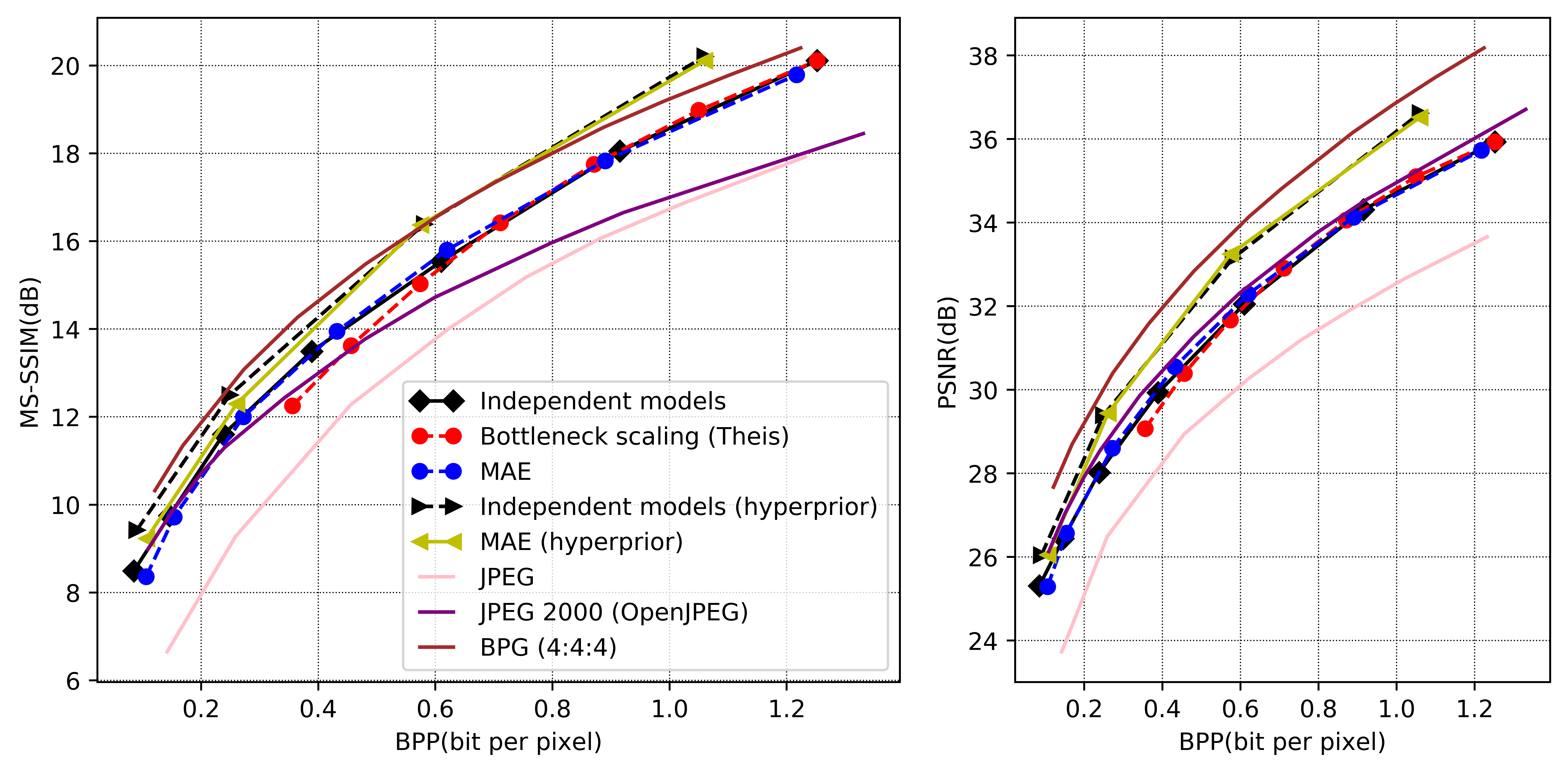}
  \caption{R-D curves for different methods on Kodak dataset.}
  \label{fig:kodak}
  \vspace{-0.4cm}
\end{figure}

\subsection{Results}
We use MS-SSIM~(dB)\footnote{MS-SSIM~(dB) is computed as $\shortminus10\log_{10}(1\shortminus\text{MS-SSIM})$~\cite{johnston2018improved}} and PSNR~(dB) to evaluate image distortion.  Fig.~\ref{fig:clic} and Fig.~\ref{fig:kodak} show the R-D operational curves for the proposed MAE and the two baselines on the CLIC 2019 Professional dataset and the Kodak dataset, respectively. In addition, it also includes JPEG~\cite{wallace1992jpeg}, JPEG2000~\cite{jpg2000} and BPG~\cite{bpg} coding methods to demonstrate their performance. For both of them, we can see that the best R-D performance is obtained by using independent models. Hyperprior models also have superior R-D performance. Bottleneck scaling is satisfactory for high bitrates, closer to the optimal R-D operational point of the autoencoder, but degrades for lower bitrates. Interestingly, bottleneck scaling cannot achieve as low bitrates as independent models since the autoencoder is optimized for a high bitrate. This can be observed in the R-D curve as a narrower range of bitrates. Note that our independent models results did not achieve the same performance as in ~\cite{balle2016end} and ~\cite{balle2018variational} due to the different training datasets.
The proposed MAEs can achieve an R-D performance very close to the corresponding independent models, demonstrating that multi-layer modulation with joint training is a more powerful mechanism to achieve  effective variable rate compression. 

The main advantage of bottleneck scaling and MAEs is that the autoencoder is shared, which results in much fewer parameters than independent models, which depend on the number of R-D tradeoffs (Table~\ref{memory_requirements}). Both methods have a small overhead due to the modulating networks or the scaling factors (which is smaller in bottleneck scaling).

\begin{figure}[!t]
  \centering
  \includegraphics[width=0.48\textwidth]{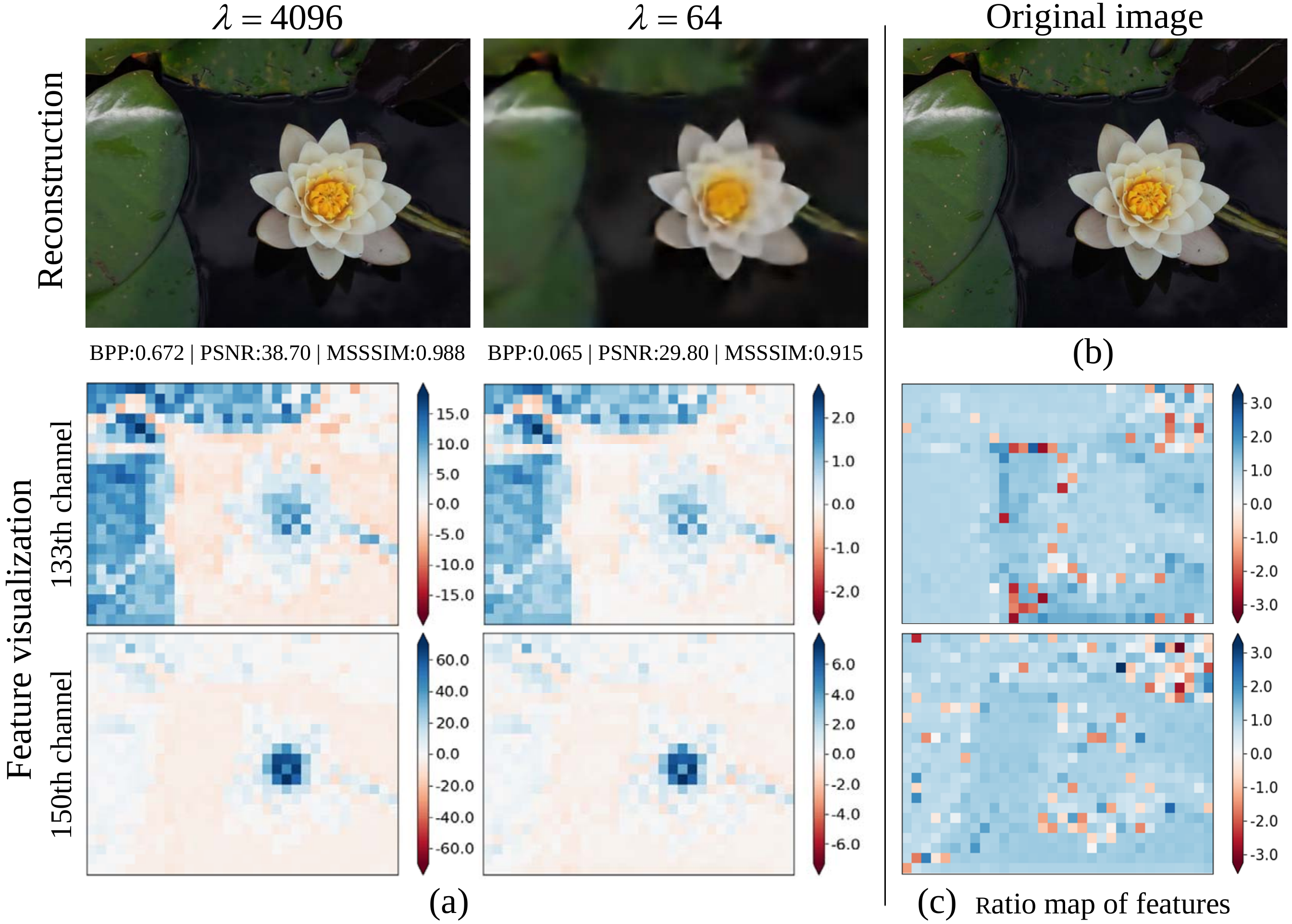}
  \caption{Modulated feature maps: (a) reconstructed images for high ($\lambda=4096$) and low ($\lambda=64$) bitrates (first row), and the corresponding feature maps for two channels of the bottleneck before quantization (2nd and 3rd rows), (b) original image for comparison, and (c) element-wise ratio (logarithmic scale) between the feature maps at the two different tradeoffs.}
  \label{example}
\end{figure}

In order to illustrate the differences between the bottleneck scaling and MAE bitrate adaptation mechanisms, we consider the image in Fig.~\ref{example}b and the reconstructions for high and low bitrates in Fig.~\ref{example}a. We show two of the 192 channels in the bottleneck feature before quantization (Fig.~\ref{example}a), and observe that the maps for the two bitrates are similar but the range is higher for $\lambda=4096$, so the quantization will be finer. This is also what we would expect in bottleneck scaling. However, a closer look highlights the difference between both methods. We also compute the element-wise ratio between the bottleneck features at $\lambda=4096$ and $\lambda=64$, and show the ratio image for the same channels of the example (Fig.~\ref{example}c). We can see that the MAE learns to perform a more complex adaptation of the features beyond simple channel-wise bottleneck scaling since different areas of the ratio map show different values (the ratio map would be uniform in bottleneck scaling), which allows MAE to allocate bits more freely when optimizing for different R-D tradeoffs, especially for low bitrates.

\begin{table}
	\caption{Model size (millions of parameters)}
	\label{memory_requirements}
	\small
	\setlength{\tabcolsep}{3pt}
	\begin{tabular}{|c|c|c|}
		\hline
		\multirow{2}{*}{Architecture}& 
		w/o hyperprior~\cite{balle2016end}& w/ hyperprior~\cite{balle2018variational} \\
		& (seven R-D tradeoffs) & (four R-D tradeoffs) \\
		\hline
		Independent models & 28.02 \textbf{M} & 40.53 \textbf{M}\\
		Bottleneck scaling~\cite{theis2017lossy} & 4.00 \textbf{M} & -\\
		Modulated AE (ours) &4.06 \textbf{M} & 10.27 \textbf{M}\\
		\hline
	\end{tabular}
	\label{tab1}
	\vspace{-0.3cm}
\end{table}
\vspace{-0.4cm}
\section{Conclusion}
In this work, we introduce the modulated autoencoder, a novel variable rate deep image compression framework, based on multi-layer feature modulation and joint learning of autoencoder parameters. MAEs can realize variable bitrate image compression with a single model, while keeping the performance close to the upper bound of independent models that require significantly more memory. We show that MAE  outperforms bottleneck scaling~\cite{theis2017lossy}, especially for low bitrates.

\bibliographystyle{IEEEtran}
\bibliography{spl}

\begin{thebibliography}{10}
\providecommand{\url}[1]{#1}
\csname url@samestyle\endcsname
\providecommand{\newblock}{\relax}
\providecommand{\bibinfo}[2]{#2}
\providecommand{\BIBentrySTDinterwordspacing}{\spaceskip=0pt\relax}
\providecommand{\BIBentryALTinterwordstretchfactor}{4}
\providecommand{\BIBentryALTinterwordspacing}{\spaceskip=\fontdimen2\font plus
\BIBentryALTinterwordstretchfactor\fontdimen3\font minus
  \fontdimen4\font\relax}
\providecommand{\BIBforeignlanguage}[2]{{%
\expandafter\ifx\csname l@#1\endcsname\relax
\typeout{** WARNING: IEEEtran.bst: No hyphenation pattern has been}%
\typeout{** loaded for the language `#1'. Using the pattern for}%
\typeout{** the default language instead.}%
\else
\language=\csname l@#1\endcsname
\fi
#2}}
\providecommand{\BIBdecl}{\relax}
\BIBdecl

\bibitem{theis2017lossy}
L.~Theis, W.~Shi, A.~Cunningham, and F.~Husz{\'a}r, ``Lossy image compression
  with compressive autoencoders,'' \emph{arXiv preprint arXiv:1703.00395},
  2017.

\bibitem{lewis1992image}
A.~S. Lewis and G.~Knowles, ``Image compression using the {2-D} wavelet
  transform,'' \emph{IEEE Transactions on Image Processing}, vol.~1, no.~2, pp.
  244--250, 1992.

\bibitem{villasenor1995wavelet}
J.~D. Villasenor, B.~Belzer, and J.~Liao, ``Wavelet filter evaluation for image
  compression,'' \emph{IEEE Transactions on Image Processing}, vol.~4, no.~8,
  pp. 1053--1060, 1995.

\bibitem{taubman2012jpeg2000}
D.~Taubman and M.~Marcellin, \emph{JPEG2000 image compression fundamentals,
  standards and practice: image compression fundamentals, standards and
  practice}.\hskip 1em plus 0.5em minus 0.4em\relax Springer Science \&
  Business Media, 2012, vol. 642.

\bibitem{shannon1948mathematical}
C.~E. Shannon, ``A mathematical theory of communication,'' \emph{Bell system
  technical journal}, vol.~27, no.~3, pp. 379--423, 1948.

\bibitem{jain1989fundamentals}
A.~K. Jain, \emph{Fundamentals of digital image processing}.\hskip 1em plus
  0.5em minus 0.4em\relax Englewood Cliffs, NJ: Prentice Hall, 1989.

\bibitem{toderici2015variable}
G.~Toderici, S.~M. O'Malley, S.~J. Hwang, D.~Vincent, D.~Minnen, S.~Baluja,
  M.~Covell, and R.~Sukthankar, ``Variable rate image compression with
  recurrent neural networks,'' \emph{arXiv preprint arXiv:1511.06085}, 2015.

\bibitem{balle2016end}
J.~Ball{\'e}, V.~Laparra, and E.~P. Simoncelli, ``End-to-end optimized image
  compression,'' \emph{arXiv preprint arXiv:1611.01704}, 2016.

\bibitem{gregor2016towards}
K.~Gregor, F.~Besse, D.~J. Rezende, I.~Danihelka, and D.~Wierstra, ``Towards
  conceptual compression,'' in \emph{Advances In Neural Information Processing
  Systems}, 2016, pp. 3549--3557.

\bibitem{toderici2017full}
G.~Toderici, D.~Vincent, N.~Johnston, S.~Jin~Hwang, D.~Minnen, J.~Shor, and
  M.~Covell, ``Full resolution image compression with recurrent neural
  networks,'' in \emph{conference on Computer Vision and Pattern Recognition},
  2017, pp. 5306--5314.

\bibitem{johnston2018improved}
N.~Johnston, D.~Vincent, D.~Minnen, M.~Covell, S.~Singh, T.~Chinen,
  S.~Jin~Hwang, J.~Shor, and G.~Toderici, ``Improved lossy image compression
  with priming and spatially adaptive bit rates for recurrent networks,'' in
  \emph{conference on Computer Vision and Pattern Recognition}, 2018, pp.
  4385--4393.

\bibitem{liu2018cnn}
D.~Liu, H.~Ma, Z.~Xiong, and F.~Wu, ``Cnn-based dct-like transform for image
  compression,'' in \emph{International Conference on Multimedia
  Modeling}.\hskip 1em plus 0.5em minus 0.4em\relax Springer, 2018, pp. 61--72.

\bibitem{mentzer2018conditional}
F.~Mentzer, E.~Agustsson, M.~Tschannen, R.~Timofte, and L.~Van~Gool,
  ``Conditional probability models for deep image compression,'' in
  \emph{conference on Computer Vision and Pattern Recognition}, 2018, pp.
  4394--4402.

\bibitem{minnen2018joint}
D.~Minnen, J.~Ball{\'e}, and G.~D. Toderici, ``Joint autoregressive and
  hierarchical priors for learned image compression,'' in \emph{Advances in
  Neural Information Processing Systems}, 2018, pp. 10\,771--10\,780.

\bibitem{li2018learning}
M.~Li, W.~Zuo, S.~Gu, D.~Zhao, and D.~Zhang, ``Learning convolutional networks
  for content-weighted image compression,'' in \emph{conference on Computer
  Vision and Pattern Recognition}, 2018, pp. 3214--3223.

\bibitem{goyal2001theoretical}
V.~K. Goyal, ``Theoretical foundations of transform coding,'' \emph{IEEE Signal
  Processing Magazine}, vol.~18, no.~5, pp. 9--21, 2001.

\bibitem{wintz1972transform}
P.~A. Wintz, ``Transform picture coding,'' \emph{Proceedings of the IEEE},
  vol.~60, no.~7, pp. 809--820, 1972.

\bibitem{rumelhart1985learning}
D.~E. Rumelhart, G.~E. Hinton, and R.~J. Williams, ``Learning internal
  representations by error propagation,'' California Univ San Diego La Jolla
  Inst for Cognitive Science, Tech. Rep., 1985.

\bibitem{balle2015density}
J.~Ball{\'e}, V.~Laparra, and E.~P. Simoncelli, ``Density modeling of images
  using a generalized normalization transformation,'' \emph{arXiv preprint
  arXiv:1511.06281}, 2015.

\bibitem{nair2010rectified}
V.~Nair and G.~E. Hinton, ``Rectified linear units improve restricted boltzmann
  machines,'' in \emph{International Conference on Machine Learning}, 2010, pp.
  807--814.

\bibitem{balle2018variational}
J.~Ball{\'e}, D.~Minnen, S.~Singh, S.~J. Hwang, and N.~Johnston, ``Variational
  image compression with a scale hyperprior,'' \emph{arXiv preprint
  arXiv:1802.01436}, 2018.

\bibitem{wallace1992jpeg}
G.~K. Wallace, ``The jpeg still picture compression standard,'' \emph{IEEE
  transactions on consumer electronics}, vol.~38, no.~1, pp. xviii--xxxiv,
  1992.

\bibitem{jpg2000}
``{JPEG2000 (OpenJPEG)},'' \url{https://pypi.org/project/Glymur/}, 2000,
  [Online; accessed 8-January-2020].

\bibitem{bpg}
F.~Bellard, ``{BPG Image Format},'' \url{http://bellard.org/bpg/}, 2014,
  [Online; accessed 8-January-2020].

\end{thebibliography}

\end{document}